\documentclass[aps,pre,showpacs]{revtex4}
\usepackage{epsf}

\begin{document}
\title{Analysis of  the $\infty$-replica symmetry 
       breaking solution of the Sherrington-Kirkpatrick model}

\author{A. Crisanti}
\affiliation{Dipartimento di Fisica, Universit\`a di Roma ``La Sapienza''}
\affiliation{Istituto Nazionale Fisica della Materia, Unit\`a di Roma, 
             P.le Aldo Moro 2, I-00185 Roma, Italy}
\email{andrea.crisanti@phys.uniroma1.it}

\author{T. Rizzo}
\affiliation{Dipartimento di Scienze Fisiche, Universit\`a ``Federico II'',
         Complesso Monte S. Angelo, I-80126 Napoli, Italy}
\email{tommaso.rizzo@inwind.it}

\date{V2.1, \today}

\begin{abstract}
In this work we analyse the Parisi's $\infty$-replica symmetry breaking 
solution of the Sherrington - Kirkpatrick model without external field
using high order perturbative expansions. 
The predictions are compared with 
those obtained from the numerical solution of the
$\infty$-replica symmetry breaking equations which are solved 
using a new pseudo-spectral code which allows for very accurate 
results. 
With this methods we are able to get more insight into the analytical
properties of the solutions. 
We are also able to 
determine numerically the end-point $x_{\rm max}$ of the plateau
of $q(x)$ and find that $\lim_{T\to 0} x_{\rm max}(T) > 0.5$.
\end{abstract}

\pacs{75.10.Nr, 75.40.Cx, 02.70.Hm}

\maketitle  	

\section{Introduction}
Since its proposal in the 80's the behaviour of the Parisi $\infty$-replica 
symmetry breaking ($\infty$-RSB) solution of the 
Sherrington-Kirkpatrick model has been extensively investigated 
both qualitatively and quantitatively \cite{MPV,FH}.
Despite this enormous amount of work, which has revealed many
of the properties of the solutions, a complete control of the solution
is still missing. One of the reasons can be traced back to
the fact that till now only low order expansions were used, moreover
applied often to reduced forms 
$\infty$-replica symmetry breaking equations valid only near the
critical temperature. From the numerical
point of view there are only few works which confirm the general 
properties of the solution but do not allow for high accuracy.
On the other hand $\infty$-replica symmetry breaking solutions
of the type encountered in the SK model have been found in
other models of interest in different fields, e.g., in computer 
science with solvability problems \cite{Luca} or in the study
of the structural glass transition \cite{G85,KT87,SNA97}. 

Motivated by these problems we believe that it would be quite useful 
to have some reliable and efficient tool to find good approximations 
of the full solution also far from the critical points. 
In this work we reconsider two approaches.
The first one is based on expansions for temperatures near the
critical temperature $T_c$. As said above previous works considered 
only low order expansions \cite{P7980,K83,S85}. 
Here, by using algebraic manipulators,
we push the expansion to rather high 
orders and resumming it via Pad\`e resummation technique we are 
able to a have good estimate of the solution for a wide range of 
temperature below $T_c$.

The second approach is numerical. Previous numerical studies 
of the $\infty$-replica symmetry breaking solution 
used a naive integration scheme based on the direct discretization
of the Parisi's equation \cite{VTP81,SDJPC84,topomoto,B90}. 
The main disadvantages of this approach
are the large amount of memory needed for a good resolution of the
solution and the numerical problems arising when $\dot{q}(x)$ is small. 
To overcome these problems we developed a new numerical scheme based on
a pseudo-spectral algorithm which allows for rather accurate 
results for all temperatures with a reasonable amount
of memory. Moreover the use of pseudo-spectral methods makes the
whole code rather fast.

We stress that while the methods we are going to discuss are applied here
to the Sherrington - Kirkpatrick model, they have a wider range
of application. In principle can be applied to any model with
$\infty$-replica symmetry breaking type solution \cite{Luca}.

We find that for the Sherrington - Kirkpatrick model the
Parisi solution $q(x)$ is not an odd function as one may expect from 
its physical meaning. At any $T < T_c$, the Taylor expansion
of $q(x)$ in powers of $x$ contains both odd as well as even powers
of $x$. The only term which is missing is $x^2$. The presence of
the fourth oder derivative was first noted by Temesvari \cite{Temes}.
Often, instead of $q(x)$, it is more useful to consider 
the overlap probability distribution function $P(q)$, 
which gives the probability of finding two states
with mutual overlap $q$ according to the Gibbs measure.
The two quantities are related by \cite{DY83,P83}:
\begin{equation}
\label{eq:pdq}
P(q)=\frac{dx}{dq}
\end{equation}
where $x(q)$ is the inverse function of $q(x)$. In the
absence of external magnetic fields the function $P(q)$ must be 
an even function of $q$. The computed function $q(x)$ is however defined
only for positive values, therefore it determines only the right branch
of the function $P(q)$.
If we define $\tilde{P}(q)=dx/dq$  for $q>0$  then 
full $P(q)$ is given by the symmetrized expression
\begin{equation}
P(q)=\frac{1}{2}\tilde{P}(-q)+\frac{1}{2}\tilde{P}(q)
\end{equation}
It is easy to see that 
the presence of non-zero even derivatives of $q(x)$ at $x=0$ makes
the function $P(q)$ non analytical at $q=0$:
\begin{equation}
P(q)=c_{0}+c_{2}q^{2}+c_{3}|q|^{3}+\ldots
\end{equation}
so that $P(q)$ has discontinuous derivatives at $q=0$.

We shall discuss two different methods of computing the expansions. 
The first, discussed in Section \ref{SecExp}, performs expansion
before imposing stationarity of the free energy functional.
The two steps however can be inverted, i.e., the expansion can be done
after stationarity is imposed, Section \ref{SecDeriv}.
The two approaches are obviously equivalent and 
the advantage of using one or the other only depends on which 
quantity one is interested in.
Since the expansions are likely to be asymptotic some resummation
scheme, such as Pad\`e discussed in Section \ref{SecPad}, are needed.
Finally in Section \ref{SecNum} we present a new integration
procedure and compare
the analytical results with those obtained from a direct numerical
solution of the $\infty$-replica symmetry breaking equations. 

\section{Expansion of the free energy functional}
\label{SecExp}
The Parisi's free energy $f$ for a the SK model in an external field
$h$ at temperature $T$ is \cite{P80}:

\begin{equation}
 -f = \frac{\beta}{4}\,\Bigl( 
                      1 - 2\,q(1) + \int_0^1dx\, q^2(x)
                         \Bigr) 
+ \int_{-\infty}^{+\infty} \frac{d y}{\sqrt{2 \pi q(0)}}
\exp\left(-\frac{(y-h)^2}{2\,q(0)}\right)\phi(0,y)
\label{eqfree}
\end{equation}
where $\phi(0,y)$ is the solution evaluated at $x=0$ 
of the the Parisi's equation
\begin{equation}
\dot\phi(x,y)=-\frac{\dot{q}(x)}{2}\,
	\Bigl[
               \phi''(x,y)+\beta\,x\,\phi'(x,y)^2 
        \Bigr]
\label{eqPhi}
\end{equation}
with the boundary condition
\begin{equation}
\phi(1,y)= \beta^{-1}\log\left(2\cosh \beta y\right)
\label{Phi1}
\end{equation}
where we 
have used the standard notation and denote derivatives with respect to 
$x$ by a dot and derivatives with respect to $y$ by a prime. 
The order parameter $q(x)$ at temperature $T$ is obtained by 
the stationarity condition of (\ref{eqfree}) with respect to variations of
$q(x)$, while the value of (\ref{eqfree})  at the stationarity point 
gives the free energy $f(T)$. 

To expand the free energy functional (\ref{eqfree}) in powers of 
$\tau = T_c - T = 1 - T$ we observe that in the absence of
external fields $q(x)$  is different from $q(1)$ only in a region 
$[0,x_{\rm max}]$ with $x_{max} = O(\tau)$ \cite{P7980}, so that 
an expansion in power of $\tau$  must correspond to an expansion 
of the same order in $x$. 
Therefore 
to compute the free energy to order $n$ we insert into eq. (\ref{eqfree})
the following expansions:
\begin{equation}
\label{eqq1}
 q(1) = \sum_{i=1}^{n-2}\, a_i\,\tau^i
\end{equation}
and
\begin{equation}
\label{eqxq}
 x(q) = \sum_{i=1}^{n-3}\,\sum_{j=0}^{n-3-i} b_{ij}\,q^i\,\tau^j.
\end{equation}
The coefficients of the expansion of the function $\phi(q,y)$ 
about $q=q(1)$ and $y=0$ can be obtained by repeated differentiation
with respect to $q$ of the equation
\begin{equation}
\frac{\partial \phi}{\partial q} = -\frac{1}{2}\left[ 
                    \frac{\partial^{2} \phi}{\partial y^{2}} 
              + x(q)\left( \frac{\partial \phi}{\partial y}\right)^{2}
                                               \right].
\end{equation}
Differentiating $j$ times with respect to $y$ this equation,
mixed derivatives $\phi^{(1,j)}(q,y)$ can be eliminated in favor of 
derivatives with respect to $y$  only.
In the absence of an external magnetic field the last term in equation
(\ref{eqfree}) reduces to $\phi(0,0)$ greatly simplifying the calculation
since at each step we can eliminate all terms containing odd 
derivatives of $\phi$ with respect to $y$, as for example
$(\partial \phi / \partial y)^{2}$ in the previous equation,
since all these vanish if evaluated at $y=0$ being $\phi(q,y)$ and even
function of $y$.

Collecting all terms with the same power of $\tau$ the free energy
functional (\ref{eqfree}) is written as
\begin{equation}
 f = \sum_{i=0}^{n}\, c_i[\{a\},\{b\}]\, \tau^i.
\end{equation}
This expression must be stationary with respect to variations of
$a$'s and $b$'s for any $\tau$. Imposing stationarity of
each $c_i$ we can find the value of the parameters $a$ and $b$.
For example to order $\tau^6$ we have: 
\begin{eqnarray}
\label{eq:qx6}
 q(x) = \left(
               \frac{1}{2} + \frac{3}{2}\,\tau + 2\,\tau^3 
               - 9\, \tau^4 + \frac{336}{5}\,\tau^5
        \right)\, x
       &+& \left(
               -\frac{1}{8} + \frac{25}{8}\,\tau + 3\,\tau^2
               + 38\,\tau^3
         \right)\, x^3
\nonumber\\
       &+& \left(
               -1 - 9\,\tau - 30\,\tau^2
         \right)\, x^4
       + \left(
               \frac{351}{320} + \frac{9189}{320}\,\tau
         \right)\, x^5
        - \frac{27}{5}\, x^6
\end{eqnarray}
and
\begin{equation}
\label{eq:xmax6}
 x_{\rm max} = 2\,\tau - 4\,\tau^2 + 12\,\tau^3 - 69\,\tau^4
               +\frac{2493}{5}\,\tau^5 - \frac{20544}{5}\,\tau^6.
\end{equation}
By using this procedure we have obtained the free energy up to order 
$30$, $q(x)$ to order $13$ and $q(1)$ to order $14$ because
despite the fact that the free energy is evaluated to order $n$, 
the variational relations allow to determine 
$x(q)$ only to order $[(n-3)/2]$ and $q(1)$ only
to order $[(n-1)/2]$.

 From eq. (\ref{eq:qx6}) we clearly see that $q(x)$ contains even powers 
of $x$, with the exclusion of $x^2$. In the next Section we shall derive
exact relations among the derivatives of $q(x)$ at $x=0$
from which follow that $q^{(2)}(x=0) = 0$ 
but $q^{(4)}(x=0) \not= 0$.


\section{Expansion of the order parameter $q(x)$}
\label{SecDeriv}

To evaluate the derivatives of 
the order parameter $q(x)$ at $x=0$ we use a variational approach
developed by Sommers and Dupont\cite{SDJPC84}. This method as also the 
advantage of leading to exact relations among derivatives of different
order, so can be used to test the findings of the previous Section
in a non-perturbative way.
The starting point is the variational form of 
the Parisi's free energy $f$:
\begin{eqnarray}
 -f &=& \frac{\beta}{4}\,\Bigl( 
                      1 - 2\,q(1) + \int_0^1dx\, q^2(x)
                         \Bigr) 
+ \int_{-\infty}^{+\infty} \frac{d y}{\sqrt{2 \pi q(0)}}
\exp\left(-\frac{(y-h)^2}{2\,q(0)}\right)\phi(0,y)
\nonumber
\\
  &-&\int_{-\infty}^{+\infty} dy\  P(1,y)\,
          \bigl[\phi(1,y)-T \log\left(2\cosh \beta y\right)\bigr]
\nonumber
\\
&+&\int_0^1dx\int_{-\infty}^{+\infty} dy\ P(x,y) 
                 \left[\dot\phi(x,y)+
                 \frac{\dot{q}(x)}{2}\,
	\Bigl[
               \phi''(x,y)+\beta\,x\,\phi'(x,y)^2 
        \Bigr]
\right].
\label{eqfrev}
\end{eqnarray}
Imposing stationarity with respect to variations of
$P(x,y)$, $P(1,y)$, $\phi(x,y)$, $\phi(0,y)$ and $q(x)$, one
obtains the variational equations:
\begin{equation}
\label{SP1}
q(x)=\int dy\, P(x,y)\,m^{2}(x,y)
\end{equation}
\begin{equation}
 \dot m(x,y)=-\frac{\dot{q}(x)}{2}\, 
           \Bigl[
          m''(x,y) + 2\,\beta\,x\, m(x,y)\ m'(x,y) 
           \Bigr]
\label{SP2}
\end{equation}

\begin{equation}
\label{SP3}
\dot{P}(x,y) = \frac{\dot{q}(x)}{2}
                   \Bigl[ 
         P''(x,y) - 2\,\beta\,x\,[m(x,y)\,P(x,y)]'
                   \Bigr]
\end{equation}
with initial conditions (in the absence of a magnetic field) 
\begin{eqnarray}
m(1,y) & = & \tanh (y/T)\label{condm} \\
P(0,y) & = & \delta (y)\label{condP}
\end{eqnarray}
These equations are the starting
point of both the expansion discussed in this Section and
the numerical solution.

A time scale $\tau_x$ can be 
associated to the order parameter $q(x)$ 
such that for times of order $\tau_x$ states with an overlap 
equal to $q(x)$ or greater can be reached by the system.
In this picture the $P(x,y)$ and $m(x,y)$ become respectively 
the probability distribution of frozen local fields $y$ 
and the local magnetization in a local field $y$ 
at the time scale labeled by $x$  \cite{SDJPC84,MPV}.

The derivatives of $q(x)$ can be obtained by successive
$x$-derivation of eq. (\ref{SP1}). The procedure is simplified 
by the use of the following identity \cite{S85}:
\begin{equation}
\label{int1}
\frac{d}{dx}\int dy\,P(x,y)\,f(x,y)=\int dy\,P(x,y)\,\Omega(x,y)\, f(x,y)
\end{equation}
where
\begin{equation}
\Omega(x,y) =\frac{\partial }{\partial x}+\frac{\dot{q}}{2}
       \left( \frac{\partial ^{2}}{\partial y^{2}} + 2\,\beta\,x\,m(x,y)
              \frac{\partial }{\partial y}\right)
\end{equation}
The application of the operator $\Omega(x,y)$
generates derivatives of the function $m(x,y)$ with respect to
$x$ and $y$. Mixed derivatives such as  $m^{(1,j)}(x,y)$
can be eliminated in favor of derivatives of $m(x,y)$ respect only 
to $y$
by deriving equation (\ref{SP2}) $j$ times with
respect to $y$.

The first application of this procedure 
yields
\begin{equation}
\dot{q}(x)=\dot{q}(x)\,\int dyP(x,y)(m')^{2}
\end{equation}
which for $\dot{q}(x)\neq 0$ simply reads
\begin{equation}
\label{d1}
1=\int dy\,P(x,y)\, m'(x,y)^{2}.
\end{equation}
The procedure can be iterated infinitely. For example, 
the next three applications leads respectively to
\begin{equation}
\label{d2}
0=-\frac{2x}{T}\int dyP(m')^{3}+\int dyP(m'')^{2}
\end{equation}
\begin{equation}
\label{d3}
\frac{2}{T}\int dyP(m')^{3}=\dot{q}\int Pdy\left( (m''')^{2}-\frac{12x}{T}m'(m'')^{2}+6\left( \frac{x}{T}\right) ^{2}(m')^{4}\right) 
\end{equation}
and
\[
\int Pdy\left( \frac{(18x\dot{q}+6x^{2}\ddot{q})(m')^{4}}{T^{2}}-\frac{(18\dot{q}T+12x\ddot{q}T-120m'x^{2}\dot{q}^{2})m'(m'')^{2}}{T^{2}}\right.
\]
\begin{equation}
\label{dev2x}
\left. +\frac{-30x\dot{q}^{2}(m'')^{2}+\ddot{q}m'''T-20x\dot{q}^{2}m'(m''')}{T}m'''-\frac{24x^{3}\dot{q}^{2}(m')^{5}}{T^{3}}+\dot{q}^{2}(m'''')^{2}\right) =0
\end{equation}
We are interested into the derivatives of $q(x)$ at $x=0$, so we 
take the limit $x\rightarrow 0$ of the above equations.
The limit can be done in trivial way and, since the function
$P(0,y)$ reduce to a $\delta$-function [see eq. (\ref{condP})],
the equations are greatly simplified.
Moreover since $m(x,y)$ is an odd function of $y$ for any
$x$ clearly $m^{(0,j)}(0,0)=0$ for any even  $j$.
In this limit equations (\ref{d1}), (\ref{d3}) and (\ref{dev2x}) 
reduce respectively
to:
\begin{eqnarray}
\label{dev10}
1 & = & m'(0,0)\label{limz2} \\
\label{dev20}
\frac{2}{T}m'(0,0)^{3} & = & \dot{q}(0)m'''(0,0)^{2}\label{limz} \\
\label{eq:d40}
\ddot{q}(0)m'''(0,0)^{2} &=& 0 
\end{eqnarray}
while equations (\ref{SP1}) and (\ref{d2}) become trivial identities.

 From equations (\ref{limz2}) and (\ref{limz}) we have 
\begin{equation}
\label{m30}
m'''(0,0)=-\sqrt{\frac{2}{T\dot{q}(0)}}\neq 0
\end{equation}
therefore (\ref{eq:d40}) implies that $\ddot{q}(0)=0$ as
already found in Ref. \cite{S85}.

To obtain information on the fourth derivative of $q(x)$ the above 
procedure must be iterated two more times. Since successive
derivatives yields expressions with a rapidly
growing number of terms we only report the limit $x\rightarrow 0$
result:
\begin{equation}
\label{dev30}
\frac{18\dot{q}(0)}{T^{2}}+q^{(3)}(0)m'''(0,0)^{2}-\frac{38\dot{q}(0)^{2}m'''(0,0)^{2}}{T}+\dot{q}(0)^{3}m^{(0,5)}(0,0)=0
\end{equation}
\begin{equation}
\label{eq:dev40}
q^{(4)}(0)m'''(0,0)-\frac{96\dot{q}(0)m'''(0,0)^{3}}{T}=0
\end{equation}
where 
equation (\ref{limz2}) and the exact result \( \ddot{q}(0)=0 \)
have been used.
Note that equation (\ref{eq:dev40}), with equation (\ref{m30}),
gives a complete determination of the quartic derivative of $q(x)$
at $x=0$  as a function of the temperature $T$ and of the first derivative
$\dot{q}(x=0)$:
\begin{equation}
q^{(4)}(0)=-\frac{96\sqrt{2}\dot{q}(0)^{5/2}}{T^{3/2}}
\end{equation}
This relation shows that the function $q(x)$ does not have a well defined
parity \cite{Temes}. 

Going to higher orders one can show that all the
even derivatives can be expressed in terms of the odd ones; for instance
we have
\begin{equation}
q^{(6)}(0)=-\frac{34272\sqrt{2}\dot{q}(0)^{7/2}}{T^{5/2}}
           -\frac{1056\sqrt{2}\dot{q}(0)^{3/2}q^{(3)}(0)}{T^{3/2}}
\end{equation}
and so on.

In the limit $T\to 0$ we have $T\,\dot{q}(0) = 0.743\pm 0.002$.
Note that if we take $\dot{q}(0)\sim 1 /T$  for $T\to 0$ the previous equations
implies that all the derivatives diverge with the temperature as 
$q^{(n)}(0)\sim 1/T^{n}$, in agreement with the Parisi-Toulouse scaling
$q(x,T)= q(\beta x)$ \cite{VTP81,PT80}. 
Note that we have derived this scaling under
strong hypothesis that it must
be valid asymptotically for $T\to 0$ and $\beta x\to 0$.

This approach also provides an alternative method to compute
the expansion of $q(x)$ in powers of $x$  and $\tau$: 
starting from $q(x)$ evaluated 
at a given order in $x$ and $\tau$
we can compute \( m^{(0,j)}(0,0) \) through (\ref{SP2})
and then $q(x)$ at the next order through the set of equations 
(\ref{limz2}),(\ref{eq:d40}), (\ref{dev30}),(\ref{eq:dev40}) and so on.
The set of equations can be solved iteratively.
By this method we were able to compute
the series expansion of $q(x)$ up to order $20$, improving the results of 
previous section.

\section{Resummation of the expansions}
\label{SecPad}

Unfortunately all the expansions derived in the previous Sections are likely to
be asymptotic and to obtain sensible estimates of
the various quantities of interest some form of resummation must be done.
Here we shall consider the standard Pad\`e approximants which
for a series of degree $N+M$ reads \cite{BO78}:
\begin{equation}
 P^N_M(x) = \frac{\sum_{i=0}^{N}\, a_i\, x^i}
                      {1 + \sum_{i=1}^{M}\, b_i\, x^i} 
\end{equation}
where the coefficients are chosen so that the first $(N+M+1)$ terms
of the Taylor expansion of $P^N_M(x)$ match the the first $(N+M+1)$
terms of the of the original series.
In the following we shall call this the Pad\`e approximant $(N,M)$.

\begin{figure}[hbt]
\begin{center}
\epsfbox{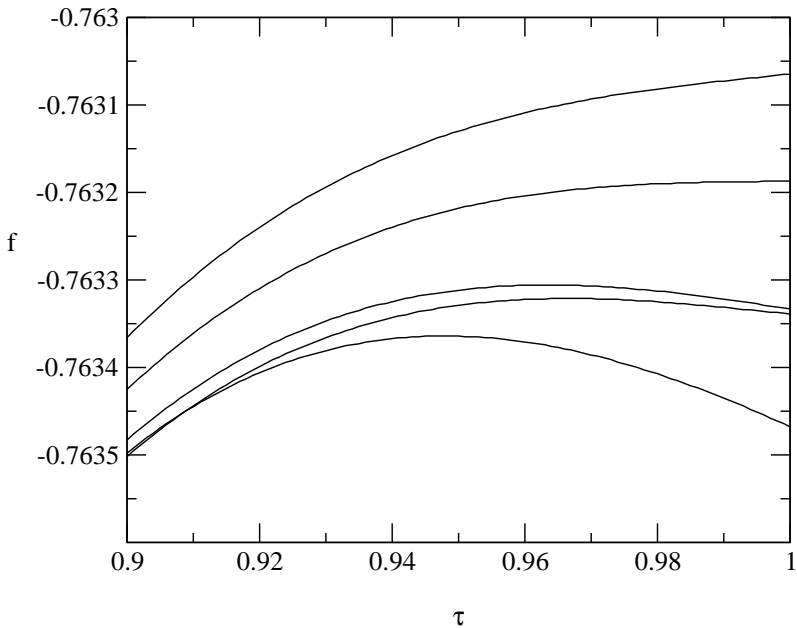}
\end{center}
\caption{Free energy as function of $\tau=1-T$ for different
         Pad\`e approximants. Top to bottom:
         $(14,16)$, $(13,12)$, $(12,11)$, $(10,11)$ and $(17,10)$.
        }
\label{fig:freepad}
\end{figure}

The first problem we faced is that despite the fact that
the series have alternate signs, they are not
Stijlties integral and therefore we cannot obtain in a systematic way
a sequence of lower and upper bounds \cite{BO78}. 
This difficulty can be overcome by noticing that
most of the quantities we are interested in, such as for example
free energy or entropy or $q(1)$,  do have a
null derivative at $T=0$. Therefore an indication on the
quality of the approximants can be obtained by analyzing the
behaviour near $T=0$. For example, the
free energy as a function of $T$ is reproduced quite well by many 
Pad\`e approximants, even at very low orders, however some of these
have a  positive derivative at $T=0$  while others negative,
see Fig. \ref{fig:freepad}.
By inspecting the figure we can safely assume that approximants with positive
derivative give an upper bound, while those with negative derivative 
a lower bound, for the true free energy \cite{Note}. 

\begin{figure}[hbt]
\begin{center}
\epsfbox{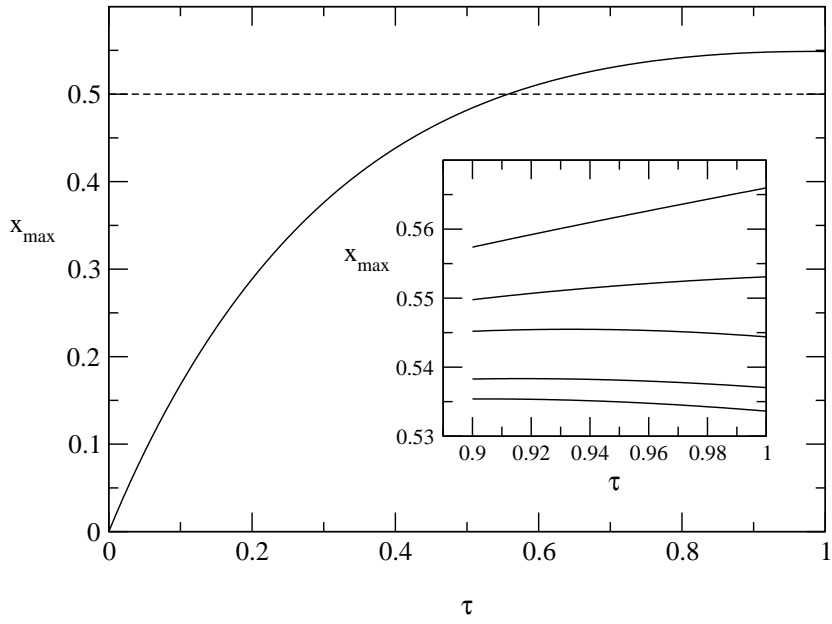}
\end{center}
\caption{$x_{\rm max}$ as function of $\tau=1-T$. 
         Inset: $x_{\rm max}$ as function of $\tau=1-T$ with different
         Pad\`e approximats. From top to bottom $(7,10)$, $(9,9)$, $(8,12)$,
         $(6,7)$ and $(7,8)$.
        }
\label{fig:xmaxpad}
\end{figure}

As a general fact we obtain that
the best Pad\`e approximants at a given order in $\tau$ 
are those with nearly
the same degree of the numerator and the denominator.
We stress, however, that as usual with asymptotic expansion an 
increase of the order in $\tau$ does not necessarily correspond to an 
improvement of the precision.
With this procedure we obtain for the free energy an estimate with at least
$16$ digits precision at $T=0.9$ and $8$ digits at $T=0.5$, and  for the 
ground state energy  $E_{0}=-.76321\pm .00003$
in agreement with Parisi's estimate $E_{0}=-.7633\pm .0001$ \cite{P80}.
A similar analysis can be used to determine the value of 
$x_{\rm max}$ as a function of temperature, the result is shown
in Fig. \ref{fig:xmaxpad}.
The value of the breaking point is finite in the limit $T\rightarrow 0$
\begin{equation}
\label{xmax}
x_{max}(0)=.548\pm .005,
\end{equation}
see inset Figure \ref{fig:xmaxpad}, and 
slightly larger than the value $1/2$ predicted by the Parisi-Toulouse scaling,
in agreement with the approximate nature of this relation \cite{VTP81,PT80}.

\begin{figure}[hbt]
\begin{center}
\epsfbox{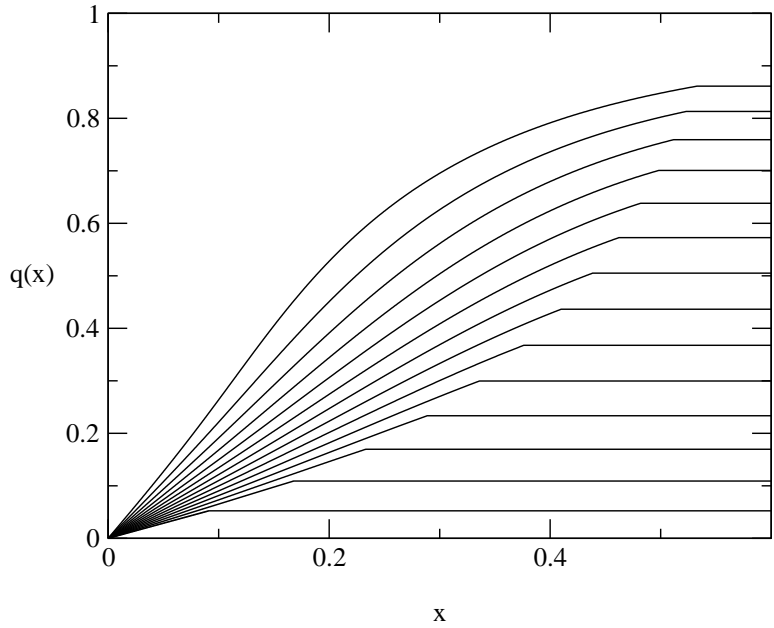}
\end{center}
\caption{$q(x)$ as function of $x$ for various temperatures.
          From bottom to top $T=0.95$ to $T=0.30$ in step of
         $0.05$.
        }
\label{fig:qxpad}
\end{figure}

The analysis of the function $q(x,T)$ is more complex, because not only
the Taylor expansion of $q(x)$ in powers of $x$ around any
$0<x<x_{max}$ is likely to be asymptotic for any fixed temperature, but 
the expansion in $\tau$ of the coefficients of the $x$-expansion 
are themselves non convergent. 
Therefore one should use a double Pad\`e expansion, one for the
coefficients and one for $q(x)$.  
The procedure however is quite difficult because we do not have a 
systematic way of choosing the best approximant and, moreover,  
coefficients of higher order are known with less precision in $\tau$. 
A better approach is to construct the function $q(x)$ directly point by
point by computing $q(mx_{\rm max})$ where $m=i/n$, 
$(i=0,1,\ldots \, n)$  for fixed $n$.
For any $m$ and $T$ the quantity $q(mx_{\rm max})$ is itself a power series
in $\tau$ which can be summed up using Pad\`e approximants. 
With this procedure the function $q(x)$ can be determined for 
different $x$-resolution just changing the value of $n$, e.g., 
$n=50,100,1000$, and using the value of $x_{\rm max}$ 
previously found, see Fig. \ref{fig:xmaxpad}.
In Figure \ref{fig:qxpad} the function $q(x)$ is shown for various
temperatures $T$.

\begin{figure}[hbt]
\begin{center}
\epsfbox{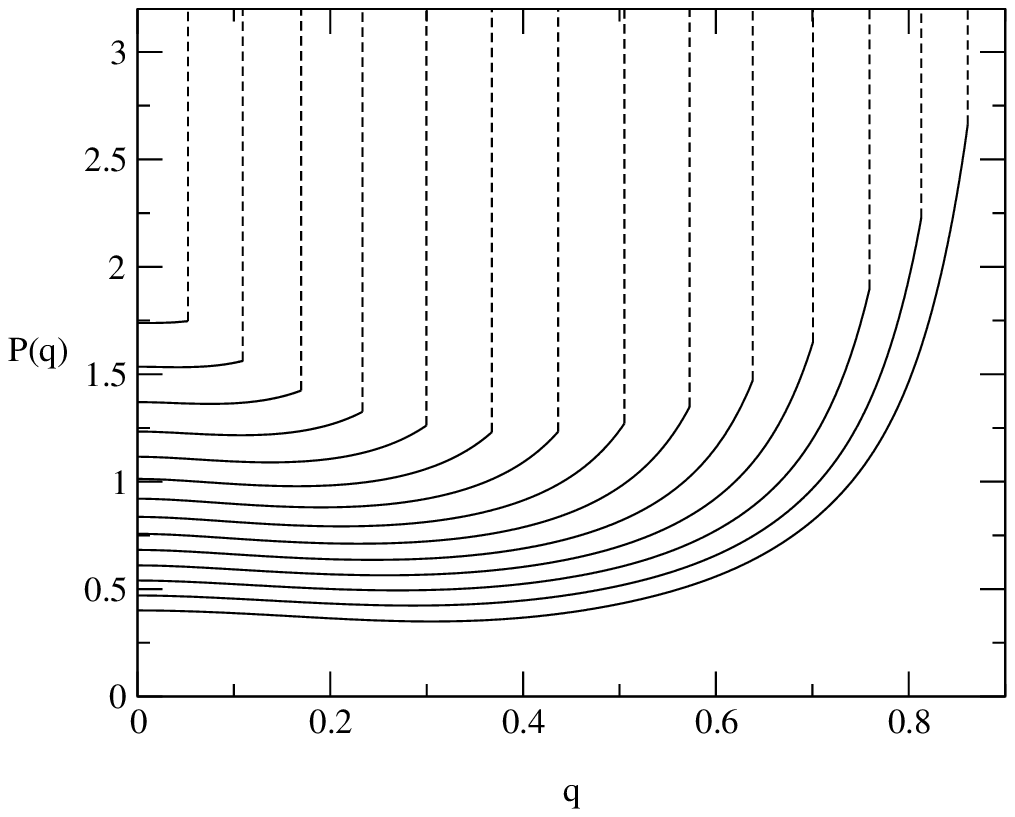}
\end{center}
\caption{$P(q)$ as function of $q$ for various temperatures.
          From left to right $T=0.95$ to $T=0.30$ in step of
         $0.05$. The data are obtained with a $(12,7)$ approximant.
         Note that $P(q)$ attains its minimum value for $q>0$. 
         This happens for any temperature
         $T<0.961938...$.
        }
\label{fig:pdqpad}
\end{figure}

This method can be extended to any function of $x$ or $q$, for example,
we computed the overlap probability function
$P(mq_{\rm max})$  in a wide range of temperature $T>0.3$, see
Figure \ref{fig:pdqpad}. We found that the best Pad\`e approximant 
is given by the $(12,7)$.
By using the relation (\ref{eq:pdq}) we can 
have an independent estimation of $q(x)$ with which to test
the precision of our results. By using a norm
$d_\infty(q,q') = \max_{0\leq x \leq 1} |q(x) - q'(x)|$ 
and expansions up to order $20$
we find, for example,
that $d_\infty(q,q') = O(10^{-5})$ for $T=0.6$ and 
$d_\infty(q,q') = O(10^{-4})$ for $T=0.4$.

\begin{figure}[hbt]
\begin{center}
\epsfbox{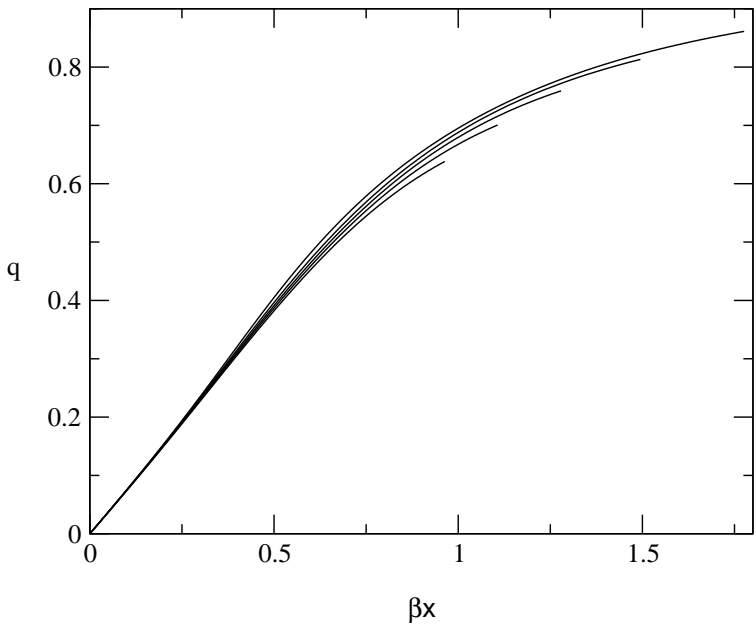}
\end{center}
\caption{$q$ as a function of $\beta x$ for different 
         values of $T$. Top to bottom 
         $T=0.30$, $T=0.35$, $T=0.40$, $T=0.45$ and $T=0.50$.
        }
\label{fig:rqxpad}
\end{figure}

The form of the function $q(x)$ confirms
the prediction of Ref. \cite{VTP81} obtained from interpolation of the $11$-RSB
solution.  In particular it confirms the approximate scaling 
$q(x,T)\sim q(x/T)$ at low temperatures, see Figure \ref{fig:rqxpad}.
Note that the scaling fails when $\beta x\sim O(1)$, in agreement
with the findings of previous section.

Finally we mention that an alternative resummation technique based 
on the Borel transform give results consistent with those
obtained with the Pad\`e approximants.


\section{Numerical Integration of the $\infty$-RSB equations}
\label{SecNum}
To check the analytical results of the previous sections we have
solved numerically the $\infty$-RSB equations (\ref{SP1}) - (\ref{condP})
on a discrete set of points in the infinite 
strip $[0\leq x\leq 1$; $-\infty< y <\infty]$
and determined $q(x)$, $P(x,y)$ and $m(x,y)$.
The numerical method is based on the iterative procedure of
Ref. \cite{topomoto}:
from an initial guess of $q(x)$ the fields $m(x,y)$,  $P(x,y)$ and the
associated $q(x)$ are computed in order as:
\begin{enumerate}
\item 
	Compute $m(x,y)$ integrating from $x=x_0$ to $x=0$ eqs. 
	(\ref{SP2}) with initial condition (\ref{condm}).
\item
	Compute $P(x,y)$ integrating from $x=0$ to $x=x_0$ eqs. 
	(\ref{SP3}) with initial condition (\ref{condP}).
\item
	Compute $q(x)$ using eq. (\ref{SP1}).
\end{enumerate}
where $x_0\leq 1$ (See later).
The steps $1.\,\to\, 2.\,\to\, 3.$ are repeated until a reasonable convergence
is reached, typically mean square error on $q$, $P$ and $m$ of the 
order $O(10^{-6})$.

The core of the numerical scheme is the integration of the partial 
differential equations (\ref{SP2}) and (\ref{SP3}) 
along the $x$ direction which, at difference with previous
numerical studies \cite{topomoto,B90}, is done
in the Fourier Space of the $y$ variables where 
the equations take the form:
\begin{equation}
\label{eq:mk}
 \frac{\partial}{\partial x}\,m(x,k) =
          \frac{k^2\dot{q}(x)}{2}\, m(x,k)
         -\frac{\beta \dot{q}(x)}{2}\, \mbox{i}k\, {\cal FT}\left[
              m^2\right](x,k)
\end{equation}
and
\begin{equation}
\label{eq:mP}
 \frac{\partial}{\partial x}\,P(x,k) =
         -\frac{k^2\dot{q}(x)}{2}\, P(x,k)
         -\beta \dot{q}(x)\, \mbox{i}k\, {\cal FT}\left[
              P\,m\right](x,k)
\end{equation}
For each wave-vector $k$ these are ordinary differential equations 
which can be integrated using standard methods.  
To avoid the time consuming calculation of the convolutions in the 
non-linear term we use a pseudo-spectral\cite{Orsz} code
on a grid mesh of $N_x\,\times\, N_y$ points,
which covers the $x$-interval $[0,x_0]$ and the $y$-interval 
$[-y_{\rm max},y_{\rm max}]$. The truncation of wave-number may 
introduce anisotropic effects for large $k$, therefore to
ensure a better isotropy of numerical treatment we perform 
de-aliasing via a $N_y/2$ truncation \cite{deal}.
Finally the $x$ integration has been performed using an 
third-order Adam-Bashfort scheme which has the advantage of reducing
the number of Fast Fourier calls \cite{AB}. 
Typical values used are $N_x=100 \div 5000$,
$N_y = 512 \div 4096$ and $y_{\rm max} = 12 \div 48$. 
The difference between the values used for $N_x$ and $N_y$ 
follows from the observation that if the solution in the $y$-direction
is smooth enough, then only few low wave-vectors are exited.
The value of the parameter $y_{\rm max}$ fixes the $y$-range 
where the solution is assumed different form zero, since in the
numerical algorithm is implicitly assumed that
\begin{equation}
 P(x,y)\equiv m(x,y) = 0 \qquad |y|> y_{\rm max}.
\end{equation}
This explain the rather large value used.  
The number of iterations
necessary to reach a mean square error on $q$, $P$ and $m$ 
of  order $O(10^{-6})$ depends on the initial 
guess of $q(x)$ but it is typically of few hundreds.

\begin{figure}[hbt]
\begin{center}
\epsfbox{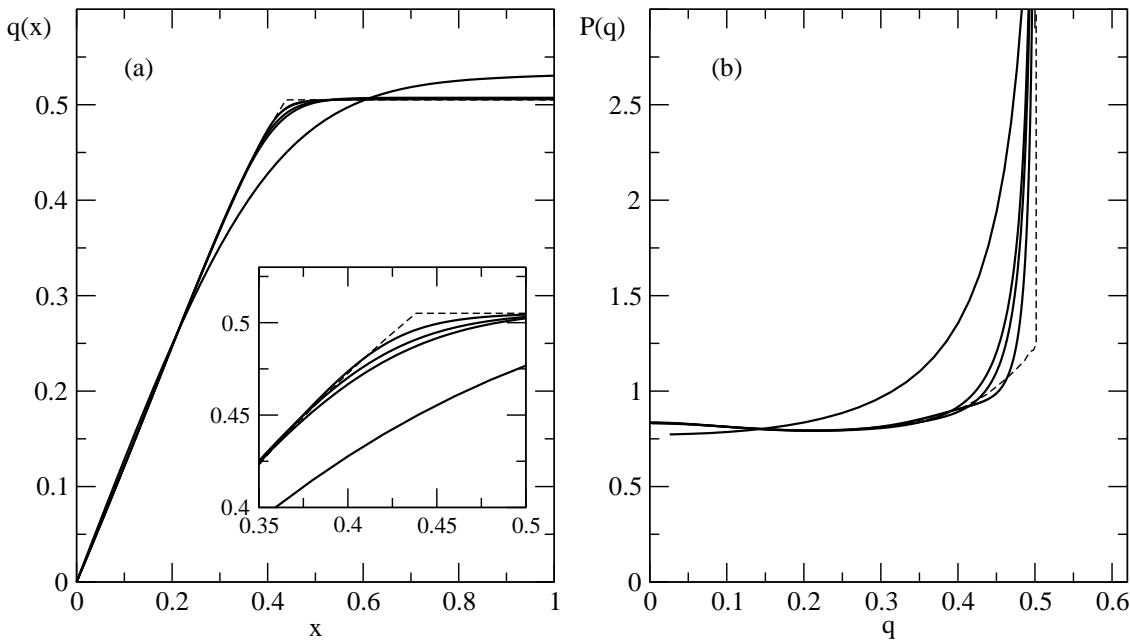}
\end{center}
\caption{$q(x)$ as function of x (panel a) and 
         $P(q)$ as function of x (panel b) at $T=0.6$ 
         for (a: bottom to top, b: left to right)
         $N_x = 50$, $N_x = 500$, $N_x = 1000$ and $N_x = 5000$.
         In all cases $x_0=1$, $y_{\rm max}=48$ and $N_y=4096$.
	 The dash line is the result from the perturbative solution
         discussed in the previous sections.
         Inset panel a: enlargement of the region near $x_{\rm max}$.
        }
\label{fig:qx-pdq.T0.6-1.000}
\end{figure}

In Figure \ref{fig:qx-pdq.T0.6-1.000}
are shown the order parameter $q(x)$ and the overlap probability
distribution function $P(q)$ at $T=0.6$ computed for 
increasing $x$-resolution and $x_0=1$. 
As expected the agreement between the 
numerical and the perturbative solutions increases with the number of 
$N_x$ of $x$-grid points. However, the convergence is not uniform:
it is rather fast far from $x_{\rm max}$ 
and much slower for $x\simeq x_{\rm max}$, see the inset of Figure 
\ref{fig:qx-pdq.T0.6-1.000} panel (a).
This is not unexpected because for $x=x_{\rm max}$ the derivative 
of the order parameter $q^{(1)}(x)$ has a cusp:
\begin{equation}
\label{eq:cusp}
\lim_{x\to x_{\rm max}^-}  q^{(1)}(x) > 0, \qquad
\lim_{x\to x_{\rm max}^+}  q^{(1)}(x) = 0
\end{equation}
making the convergence more difficult. We recall that in deriving
equations (\ref{SP2}) and (\ref{SP3}) differentiability of $q(x)$ was assumed.
The use of lower order integration schemes, as 
second-order Adam-Bashfort or Euler schemes, does not give sensible 
improvements.

\begin{figure}[hbt]
\begin{center}
\epsfbox{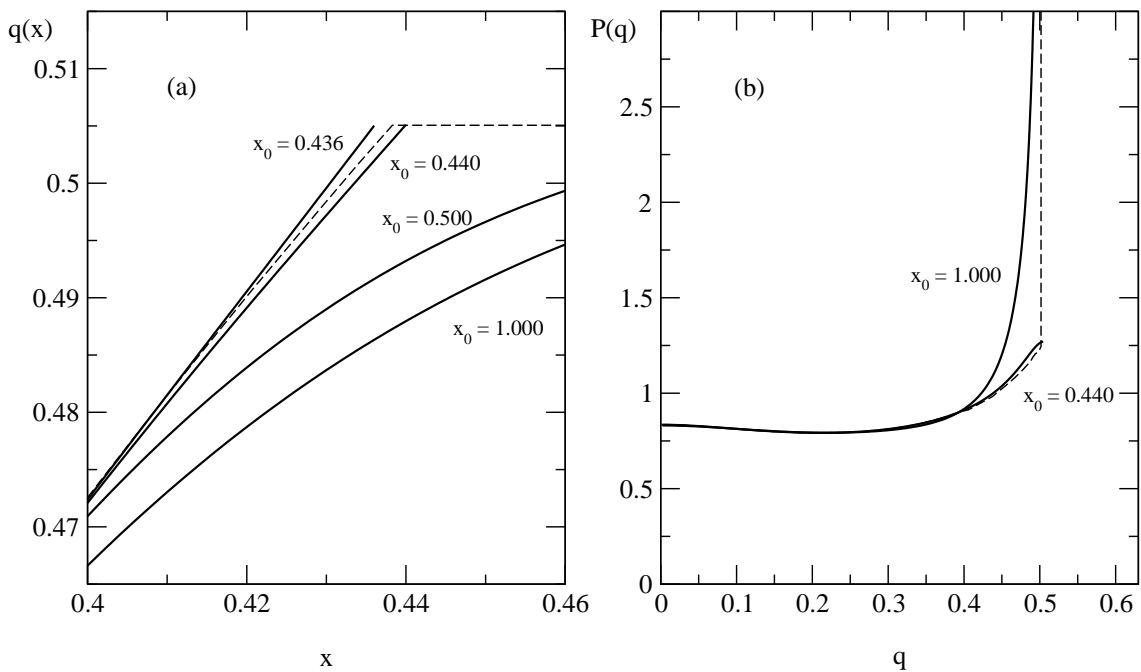}
\end{center}
\caption{Panel a: $q(x)$ for $T=0.6$ near $x_{\rm max}$ for different $x_0$.
         Panel b: $P(q)$ for $T=0.6$ for different $x_0$ .
         In all cases $N_x = 500$, $y_{\rm max}=48$ and $N_y=4096$.
	 The dash line is the result from the perturbative solution
         discussed in the previous sections.
        }
\label{fig:qx-pdq.T0.6-x0}
\end{figure}

Larger values of $N_x$ requires larger needs of computer
memory therefore to increase the precision we adopted a different approach. 
Since $\dot{q}(x)=0$ for $x>x_{\rm max}$  equations (\ref{SP2}) and (\ref{SP3})
are trivial in this range and we can reduce the upper bound of the 
$x$-integration from $x=1$ to $x=x_0=x_{\rm max}$. This obviously requires
the knowledge of $x_{\rm max}$ for the given temperature. However if we 
assume {\it no a priori} knowledge of $x_{\rm max}$  we must proceeds for
successive approximations: we start from $x_0=1$ an then reducing it until
we `hit' the value of $x_{\rm max}$. This procedure is simplified by the
fact that if $x_0<x_{\rm max}$ the shape of $q(x)$ near $x_0$ changes
dramatically with the concavity passing from negative values for
$x_0>x_{\rm max}$ to positive values for $x_0<x_{\rm max}$.
In Figure \ref{fig:qx-pdq.T0.6-x0} panel (a) are shown $q(x)$ (panel a) and
$P(q)$ (panel b) at $T=0.6$ for different values of $x_0$, 
the improvement in rather evident. 
As additional check we have considered the equality 
\begin{equation}
 1 - \int_{0}^{1} dx\, q(x) = T
\end{equation}
which is satisfied by our numerical solution for all studied temperatures 
with at least four digits. For example we for $T=0.8$ we get $0.79999(4)$,
while for $T=0.5$ the value is $0.49999(3)$.

\begin{figure}[hbt]
\begin{center}
\epsfbox{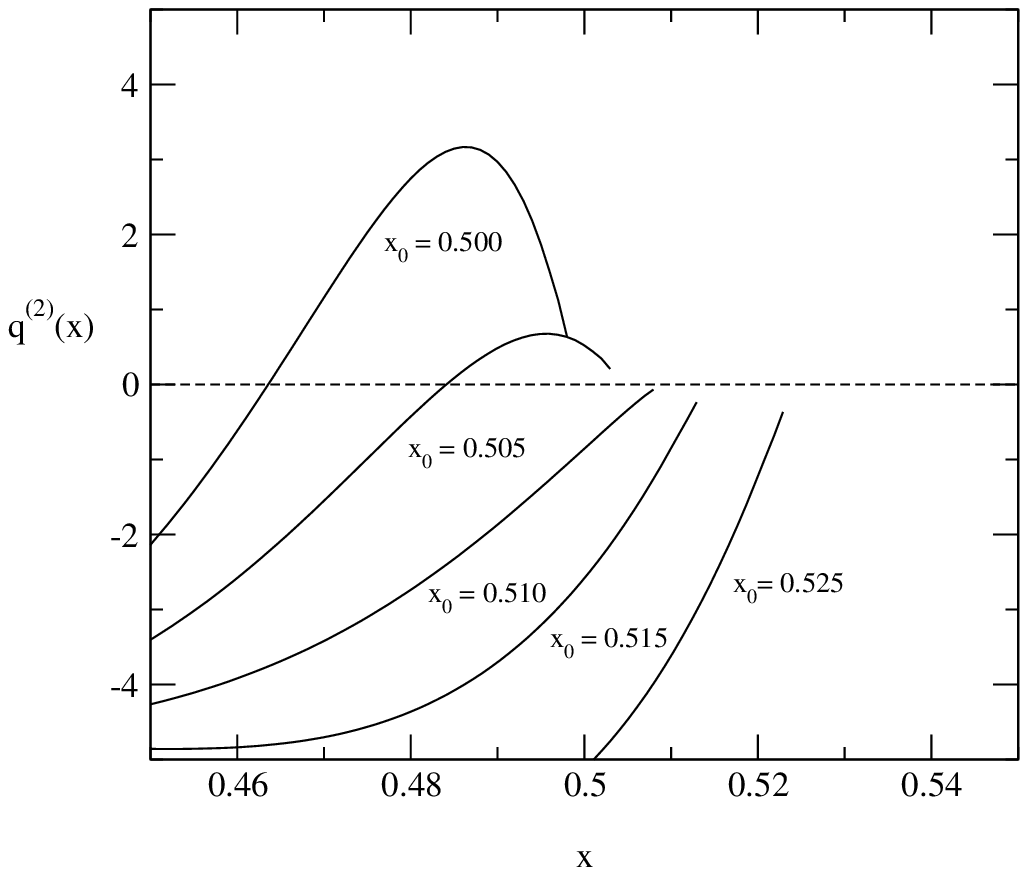}
\end{center}
\caption{Second derivative of $q(x)$ for $T=0.4$ for different $x_0$ 
         and $N_x = 500$, $y_{\rm max}=48$ and $N_y=4096$.
        }
\label{fig:q2x.T0.4-x0-500}
\end{figure}

Note that not only by fine tuning of $x_0$  
we can have a good solution for $q(x)$ at the given temperature, 
but we also have {\it the value} of $x_0$. This is best seen by 
analyzing the concavity of $q(x)$ near $x_0$. 
In Figure \ref{fig:q2x.T0.4-x0-500}
we show the second derivative of $q(x)$ near $x_0$ for $T=0.4$ and $N_x=500$,
from which one may conclude that $0.505 < x_{\rm max} < 0.510$. 

\begin{figure}[hbt]
\begin{center}
\epsfbox{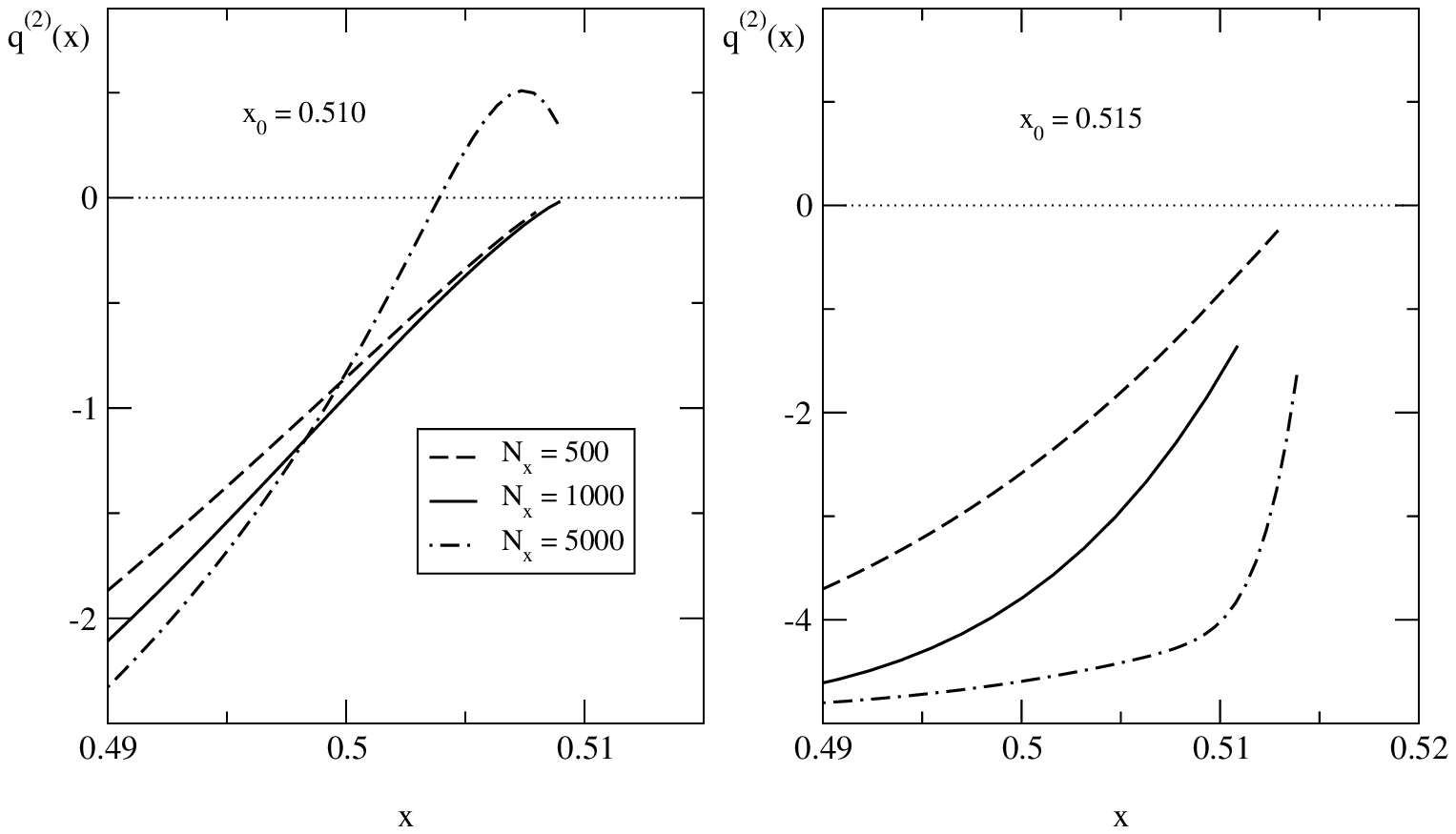}
\end{center}
\caption{Second derivative of $q(x)$ for $T=0.4$ for different $N_x$ 
         and $x_0 = 0.510$, 
         $y_{\rm max}=48$ and $N_y=4096$.
	 Left panel: $x_0 = 0.510$, 
         increasing $N_x$ leads to a positive
         value of $q^{(2)}(x_0)$ implying $x_0 = 0.510 < x_{\rm max}$.
	 Right panel: $x_0 = 0.515$, 
         increasing $N_x$ leads to a more negative 
         value of $q^{(2)}(x_0)$ implying $x_0 = 0.515 > x_{\rm max}$.
        }
\label{fig:q2x.T0.4-x0.510-0.515}
\end{figure}

A careful analysis of the stability of this results as function
of $N_x$, see Figure \ref{fig:q2x.T0.4-x0.510-0.515},
reveals, however,  
that the correct estimation is $0.510 < x_{\rm max} < 0.515$,
in rather good agreement with the analytical result 
$x_{\rm max} = 0.5111\pm 0.0002$.
The same analysis for $T=0.6$ leads to $0.438 < x_{\rm max} < 0.440$.

\begin{figure}[hbt]
\begin{center}
\epsfbox{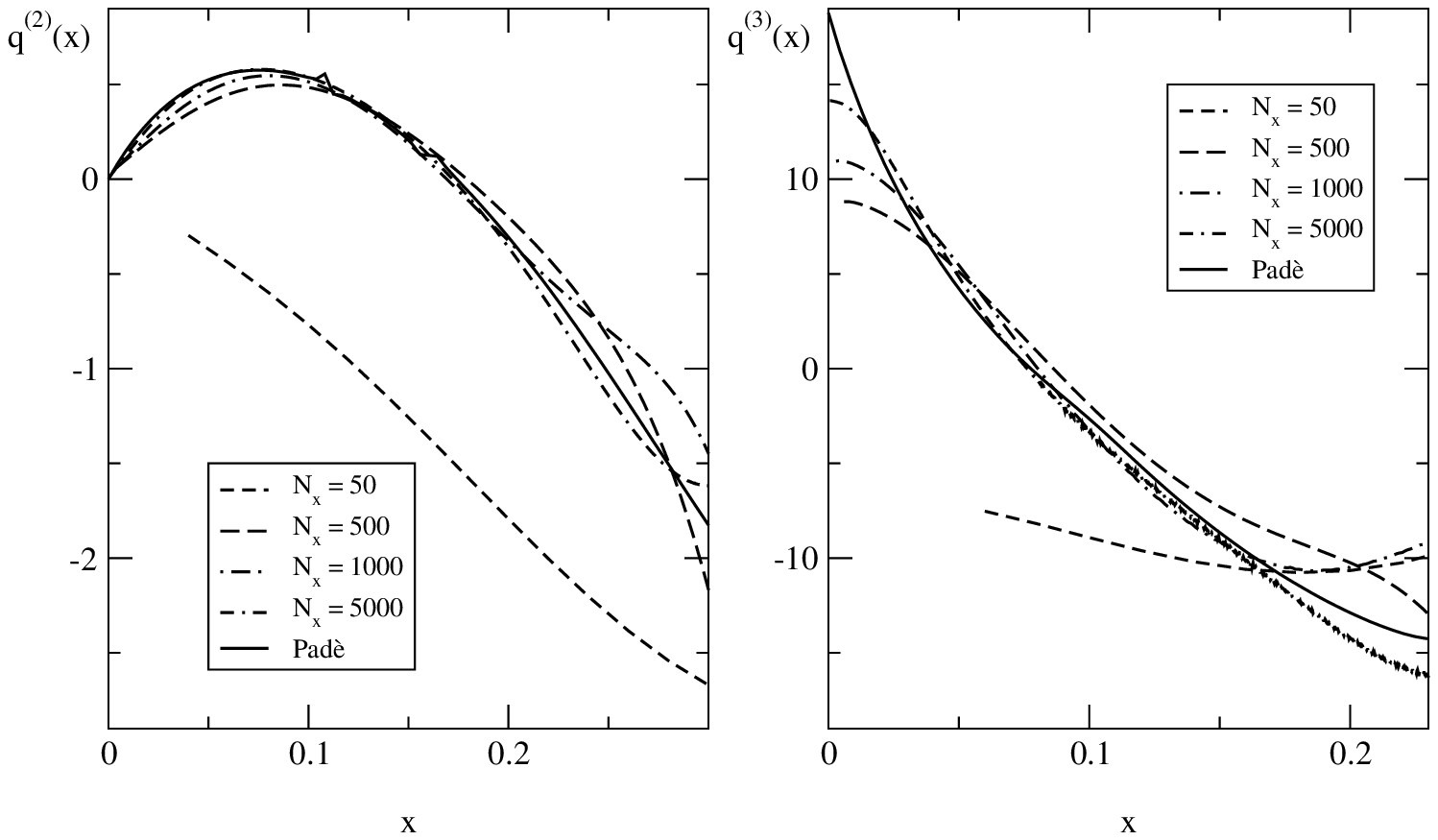}
\end{center}
\caption{Left panel: second derivative of $q(x)$ at $T=0.6$ 
         and  different $N_x$.
 	 Right panel: third derivative of $q(x)$ at $T=0.6$ 
         and  different $N_x$.
         In both cases $x_0 = 1$, $y_{\rm max}=48$ and $N_y=4096$.
         The full line is the perturbative result. 
        }
\label{fig:q2-3x.T0.6-1.000}
\end{figure}

We are now in the position of checking the results of previous section 
about the derivative of the order parameter at $x=0$, and in particular
the conclusion
\begin{equation}
\label{eq:q30}
\lim_{x\to 0} q^{(3)}(x) > 0.
\end{equation} 
In Figure \ref{fig:q2-3x.T0.6-1.000} we show the second and third
derivative of $q(x)$ obtained from numerical differentiation
of $q(x)$. The agreement with the perturbative result is sufficiently good, 
moreover from the right panel of 
Figure \ref{fig:q2-3x.T0.6-1.000}
we clearly see that the prediction (\ref{eq:q30}) is verified.

\begin{figure}[hbt]
\begin{center}
\epsfbox{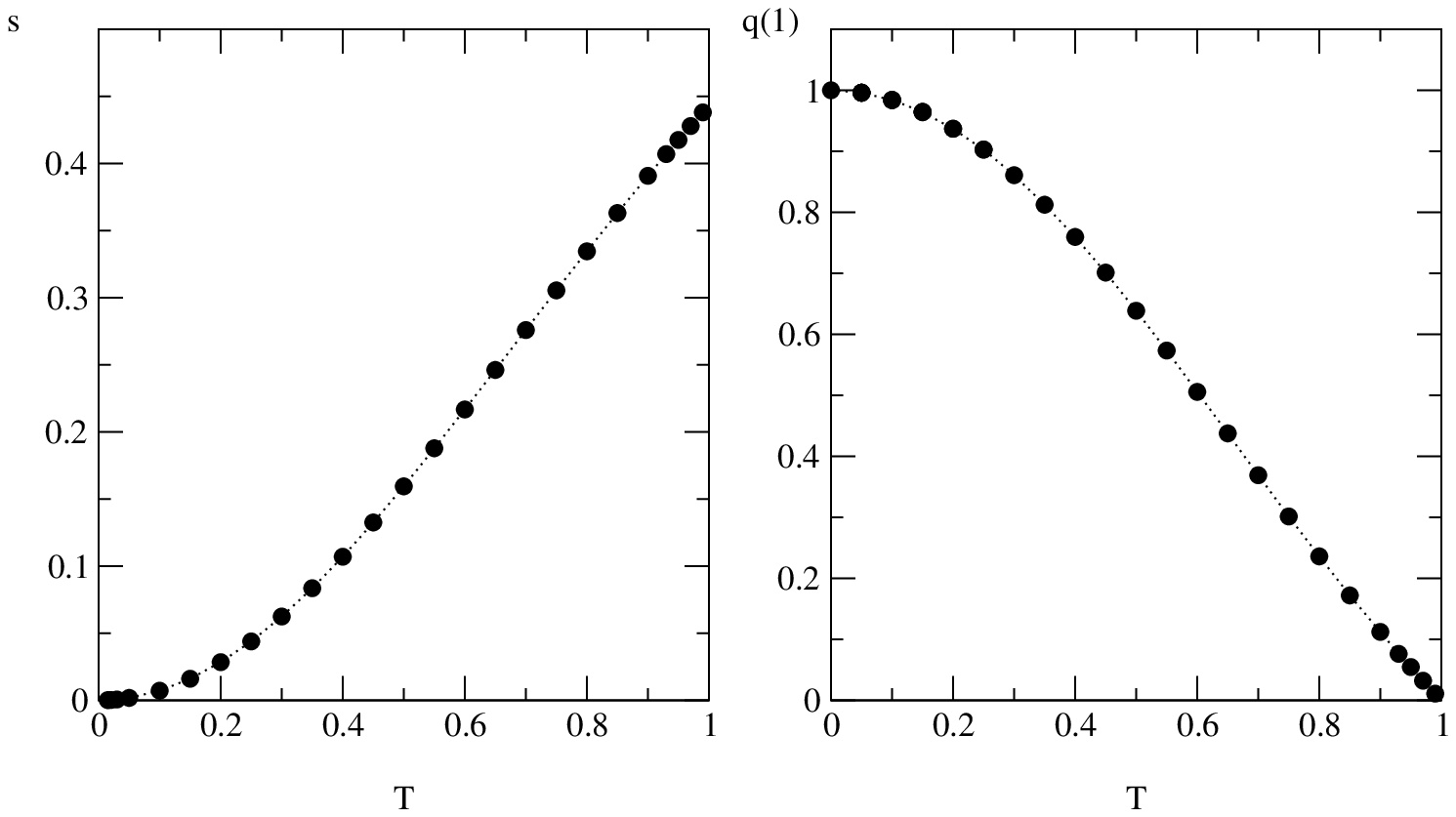}
\end{center}
\caption{Left panel: entropy $s$ as function of temperature $T$. 
         Right panel: $q(1)$ as function of temperature $T$.
        }
\label{fig:s-q1}
\end{figure}

We conclude this Section with a short discussion on the entropy which,
using the stationarity of the 
free energy functional (\ref{eqfrev}), can be written as:
\begin{equation}
 s = -\frac{\beta^2}{4}\bigl[1-q(1)\bigr]^2
     + \int_{-\infty}^{\infty} dy \ P(1,y)
\bigl[\log 2 \cosh \beta y - y \ \tanh(\beta y)\bigr].
\label{s}
\end{equation}
For other equivalent forms see, e.g., Ref. \cite{Luca}.
The entropy as function of temperature is shown in the left panel of
Figure \ref{fig:s-q1}. The entropy must vanish quadratically 
with the temperature as $T\to 0$ \cite{SDJPC84}. 
From our numerical data we find
\begin{equation}
\lim_{T\to 0} \frac{s(T)}{T^2} = a \simeq 0.72
\end{equation}
to be compared with $0.718\pm 0.004$ of the analytic expansions.

In the limit $T\to 0$ the quantity $1-q(1)$ must also vanish as $T^2$
\cite{SDJPC84}.
The behaviour of $q(1)$ as function of $T$ is shown in the right panel of
Figure \ref{fig:s-q1}. Using this data we obtain
\begin{equation}
\lim_{T\to 0} \frac{1 - q(1)}{T^2} \simeq 1.60 
\end{equation}
in very good agreement the value $1.60\pm 0.01$ obtained with the expansions
of previous sections.

\section{Conclusions}
\label{SecCon}
In this paper we have studied the properties of the $\infty$-replica symmetry
breaking solution of the Sherrington - Kirkpatrick model without external 
fields. Using high order expansions in $\tau = T_c - T$ we are able to
compute the order parameter $q(x)$ and other relevant 
quantities for a large range of temperatures with high precision. 
In particular we found that $q(x)$ {\it is not} an odd function of $x$,
confirming the prediction of Ref. \cite{Temes}.
Direct consequence of this is that the overlap probability
distribution function $P(q)$ has discontinuous derivatives at $q=0$.
Another consequence of our findings is that the Parisi-Toulouse
scaling becomes exact asymptotically for $T\to 0$ {\it and} 
$\beta x\to 0$, while for $T\to 0$ is a fairly good approximation.
This is also consistent with the $T=0$ limit of the breaking
point which we found to be $x_{max}(0)=.548\pm .005$.

Having reached very high orders we can reasonably
speculate on the analytical
properties of the function \( q(x) \). In particular we believe that 

\begin{itemize}
\item All the expansions in power of \( \tau  \) are asymptotic expansions.
\item At any temperature, the function \( q(x) \) is infinitely 
differentiable
but not analytical for any \( x \), in particular the Taylor expansion
of the function \( q(x) \) around any \( 0<x<x_{max} \) does not converge
but is asymptotic.
\end{itemize}
This singular behaviour is not connected neither with the replica
limit nor with the Parisi Ansatz, actually it originates from the
singularities in the complex plain of the initial condition of the
Parisi equation: \( f(1,y)=\ln 2\cosh \beta y \). This is clearly
seen for the replica-symmetric solution\[
q=\int ^{+\infty }_{-\infty }\frac{dz}{\sqrt{2\pi }}e^{\frac{z^{2}}{2}}\tanh ^{2}(\beta \sqrt{q}z)\]
In this case it is easy to prove that the expansion of \( (1-T^{2}) \)
in powers of \( p=\beta ^{2}q \) is asymptotic because it corresponds
to substitute \( \tanh ^{2}z \) in the integrand with its Taylor
expansion which is not convergent on the whole real axes. Then one
can prove that the expansion of \( q \) in powers of \( \tau =1-T \)
is asymptotic recalling that standard manipulation (e.g. multiplication,
division, inversion...) on an asymptotic expansion in power series
do not change its character. A detailed treatment of the RSB solution
is much more complex, but the origin of the asymptotic character is
likely to be the same. Indeed an expansion in small \( \tau  \) (and
therefore in small \( q \)) corresponds to an expansion in small
\( y \) of all the quantities like \( f(x,y) \) and \( m(x,y) \);
the appearance of integrals of the form \( \int Pfdy \)where \( P(x,y)\sim \exp (y^{2}/x) \)
generates asymptotic expansions since the Taylor expansions of \( f(x,y) \)
and \( m(x,y) \) in powers of \( y \) do not converge on the whole
real axes. These arguments can be very useful in practice to guess
the position of the singularities of the Borel transform 
if one want to sum the expansions through a conformal mapping
\cite{Pbook}.
For instance in the expression of the free energy appear integrals
of the following form:\begin{equation}
\label{as2}
\int ^{+\infty }_{-\infty }\frac{dz}{\sqrt{2\pi \tau }}e^{-\frac{z^{2}}{2\tau }}\ln \cosh (z)
\end{equation}
The singularities of the Borel transform of the previous integral
are located on a cut running from \( -\infty  \) to \( -\pi ^{2}/8 \)
and a possible guess is that this be also the singularity structure
of the Borel transform of the free energy. 
This guess is supported by the analysis of the series expansions.

The analytical results have been compared with numerical solutions
of the $\infty$-replica symmetry breaking equations. We have
developed a new numerical approach based on a pseudo-spectral code
which leads to a strong enhancement of the quality of the numerical 
results. We have also shown how, for example, to determine
the value of $x_{\rm max}$ numerically. 
In all cases the agreement between the numerical and the analytical
results is rather good.

We conclude by stressing that
our results go beyond the interest on the Sherrington - Kirkpatrick model,
since the method we used here are far more general and can be
employed to a wider class of models with generalized $\infty$-replica
symmetry breaking equations such as those introduced in Ref. \cite{Luca}.
In particular in this reference the numerical 
method was applied to the 3-SAT model, and the extension to other relevant
models is under development.


\begin{thebibliography}{99}

\bibitem{MPV} M. Mezard, G. Parisi, M.Virasoro, 
              {\it Spin Glass Theory and Beyond}, 
               World Scientific, Singapore  (1987).

\bibitem{FH} K.H. Fischer and J.A. Hertz,
       {\it Spin-Glasses}
       (Cambridge University Press, 1991)       


\bibitem{Luca} A. Crisanti, L. Leuzzi and G. Parisi,
       J. Phys. A: Math. Gen. {\bf 35}, 481 (2002).

\bibitem{G85} E. Gardner, Nucl. Phys. {\bf B257}, 747 (1985);


\bibitem{KT87} T.R. Kirkpatrick and D. Thirumalai,
               Phys. Rev. B {\bf 36}, 5388 (1987).


\bibitem{SNA97} M. Sellitto, M. Nicodemi and J.J. Arenzon
                J. Phys. I France {\bf 7}, 945 (1997).
      
\bibitem{P7980} G. Parisi, 
                Phys. Rev. Lett. {\bf 43}, 1754 (1979);
                J. Phys. A {\bf 13}, L115 (1980).

\bibitem{K83} I. Kondor,
              J. Phys. A {\bf 16}, L127 (1983).

\bibitem{S85} H.-J. Sommers, J. Physique Lett. {\bf 46}, L-779 (1985).

\bibitem{VTP81} J. Vannimenus, G. Toulouse and G. Parisi
             J. Physique {\bf 42}, 565 (1981).

\bibitem{SDJPC84} H. J. Sommers, W. Dupont, J. Phys. C {\bf{17}} (1984)
 5785-5793.

\bibitem{topomoto} K. Nemoto, J. Phys. C {\bf 20}, 1325 (1987).

\bibitem{B90} P. Biscari, 
              J. Phys.  A {\bf 23}, 3861 (1990)

\bibitem{Temes} T. Temesvari, 
	 J. Phys. A {\bf 22}, L1025 (1989).

\bibitem{DY83} C. De Dominicis and A.P. Young,
              J. Phys. A {\bf 16}, 2063 (1983);

\bibitem{P83} G. Parisi, Phys. Rev. Lett. {\bf 50}, 1946 (1983).

\bibitem{P80} G. Parisi, J. Phys. A {\bf 13}, L115 (1980).

\bibitem{PT80} G. Parisi and Toulouse,
               J. Physisque Lett. {\bf 41}, L-361 (1980).

\bibitem{BO78} C.M. Bender and S.A. Orszag,
             {\it Advanced Mathematical Methods for Scientists and
                  Engineers}, McGraw-Hill (1978).

\bibitem{Note} We note that it is possible to impose a null derivative at 
               $T=0$ directly into the Pad\`e approximants. 
               This however does not produce a
               sensible improvement of the precisions at a given order, 
               making at the same time more difficult to have a control 
               on the convergence.

\bibitem{Orsz} S. A. Orszag, {\it Studies in applied mathematics}
	(Cambridge University, Cambridge, 1971), Vol. 4, p 293.

\bibitem{deal} G. S. Patterson and S. A. Orszag, 
	Phys. Fluids {\bf 14}, 2538 (1971)

\bibitem{AB} See, e.g., J.H. Ferziger and M. Peri\'c,
             {\it Computational Methods for Fluid Dynamics},
             Springer-Verlag (1996).

\bibitem{Pbook} G. Parisi, 
                {\it Statistical Field Theory},
                Addison Wesley (1988)


\end{thebibliography}
\end{document}